\documentclass[traditabstract]{aa}

\usepackage{natbib}
\usepackage{graphicx}

\usepackage{txfonts}

\newcommand{\be}{\begin{equation}}
\newcommand{\ee}{\end{equation}}

\def\ba{\begin{eqnarray}}
\def\ea{\end{eqnarray}}

\def\msol{M_\odot}

\def\he3{^3He}

\def\ltsima{$\; \buildrel < \over \sim \;$}
\def\simlt{\lower.5ex\hbox{\ltsima}}
\def\gtsima{$\; \buildrel > \over \sim \;$}
\def\simgt{\lower.5ex\hbox{\gtsima}}

\newcommand{\cp}{\citep}
\newcommand{\ct}{\citet}

\begin{document}

\title{Local heating due to convective overshooting and the solar modelling problem}
\titlerunning{Local heating due to convective overshooting and the solar modelling problem}

\author{I. Baraffe \inst{1,2}, T. Constantino \inst{1}, J. Clarke\inst{1}, A. Le Saux \inst{1,2}, T. Goffrey\inst{4}, T. Guillet\inst{1}, J. Pratt\inst{3}, D. G. Vlaykov\inst{1}}
%\author{I. Baraffe \inst{1,2} et al.}
\authorrunning{Baraffe et al.}

\offprints{I. Baraffe} 
   
\institute{
University of Exeter, Physics and Astronomy, EX4 4QL Exeter, UK
(\email{i.baraffe@ex.ac.uk})
\and
\'Ecole Normale Sup\'erieure, Lyon, CRAL (UMR CNRS 5574), Universit\'e de Lyon, France
\and
Department of Physics and Astronomy, Georgia State University, Atlanta GA 30303, USA
\and
Centre for Fusion, Space and Astrophysics, Department of Physics, University of Warwick, Coventry, CV4 7AL, UK
}

\date{}

\abstract{Recent hydrodynamical simulations of convection in a solar-like model suggest that penetrative convective flows at the boundary of the convective envelope modify the thermal background in the overshooting layer. Based on these results, we implement in one-dimensional stellar evolution codes a simple prescription to modify the temperature gradient below the convective boundary of a solar model. This simple prescription qualitatively reproduces the behaviour found in the hydrodynamical simulations, namely a local heating and smoothing of the temperature gradient below the convective boundary.   We show that introducing local heating in the overshooting layer can reduce the sound-speed discrepancy usually reported between solar models and the structure of the Sun inferred from helioseismology. It also affects key quantities in the convective envelope, such as the density, the entropy, and the speed of sound. These effects could  help reduce the discrepancies between solar models and observed constraints based on seismic inversions of the Ledoux discriminant.  Since mixing due to overshooting and  local heating are the result of the same convective penetration process, the goal of this work is to invite solar modellers to consider both processes for a more consistent approach.
}


\maketitle

\section{Introduction}
\label{introduction}

Modelling the internal structure of the Sun is still a challenge. A recent review by \ct{christensen21} describes in detail the long-standing efforts to improve solar models.  The solar modelling problem refers to the discrepancy between helioseismology and solar interior models that adopt low metallicities predicted by the three-dimensional (3D) atmosphere models of, for example, \ct{asplund09} and \ct{caffau11}, in contrast to the  high metallicities based on previous literature compilations by, for example, \ct{anders89} and \ct{grevesse93}. 
 \ct{asplund21} have recently confirmed with state-of-the-art 3D simulations  the
relatively low metal abundances for the Sun.
 \ct{asplund21} consider that their study yields the most reliable solar abundances available today, suggesting that the solar modelling problem is no longer a problem of abundances but rather a problem of stellar physics. The treatment of mixing below the convective zone is one of the key processes that could improve solar models.
Several studies indeed reveal that the process of convective penetration, also called overshooting, at the bottom of the convective envelope could play an important role in improving the agreement between solar models and helioseismic constraints \cp[see for example][]{christensen11,zhang12,buldgen19a}. 
Overshooting in solar models has most often been treated using diffusive or instantaneous chemical mixing.  A temperature gradient that sharply transitions from a nearly adiabatic form to a radiative form is usually assumed, as suggested by the theoretical work of  \ct{zahn91}. Models with a smoother transition have also been investigated.
%inspired by semi-analytical models \cp[e.g.][]{rempel04} or numerical simulations \cp[e.g.][]{Brummel 02} of convective overshoot.
Based on the analysis of  models with different stratifications near the base of the convective zone, \ct{christensen11} found that models that better fit the helioseismic data have a weakly sub-adiabatic temperature gradient in the lower part of the convective zone and a smooth transition to the radiative gradient in the overshooting layer. But \ct{christensen11} noted that the required temperature stratification is difficult to reconcile with existing overshooting models and numerical simulations.  They concluded that only non-local turbulent convection models could produce the desired degree of smoothness in the transition \cp[see for example][]{zhangli12, zhang12}. But these non-local models remain uncertain, and their description of overshooting under the conditions found at the base of the solar convective zone is yet to be validated. \ct{zhang19} explored the impact of overshooting by introducing a parametrised turbulent kinetic energy flux based on a  model with parameters that are adjusted to improve the helioseismic properties. They suggest that amelioration can be obtained specifically below the convective envelope.  However, \ct{zhang19} find that this model cannot solve the whole solar problem because such a flux worsens the sound-speed profile in the deep radiative interior of their solar model.
Given the uncertainties regarding  the temperature stratification of the overshooting region, solar modellers have considered these effects as secondary and have focused their efforts on exploring the impact of solar abundances, microphysics (opacities, equations of state, nuclear reaction rates), and chemical mixing and diffusion \cp[see details and references in the review of][]{buldgen19b}. Additional, more exotic effects such as early disk accretion or solar-wind mass loss \cp[][]{zhang19, kunitomo21} are also attracting increasing attention.

%The motivation for adopting one of these two temperature gradients is inspired by the work of \ct{zahn91} which makes the distinction between {\it overshooting}, characterised by penetrative motions that do not alter the stable temperature gradient and {\it penetration}, characterised by penetrative flows that establish a nearly adiabatic stratification.  Solar models adopting one of these two temperature gradients have led to inconclusive results. 
To reinvigorate the debate,  \ct{buldgen19a} recently highlighted once again how the transition of the temperature gradient just below the convective envelope can significantly impact the disagreement between solar models and helioseismic constraints. Their results, based on a method that combines multiple structural inversions,  suggest that the transition in temperature gradient is improperly reproduced by adopting either an adiabatic  or a radiative temperature gradient in the overshooting layer. The solution should be somewhere in between these two extremes. \ct{christensen18} also note that
an increase in the temperature at the transition would
remove a remaining small sharp dip in the speed of
sound immediately beneath the convective zone of the model. A major difficulty is to disentangle the effects of overshoot from the effects of opacities, which can also alter the temperature gradient in these layers. Given  the large number of parameters to deal with in order to improve solar models and the current lack of strong arguments in favour of modifying the thermal stratification in the overshooting layer, there has been no real motivation to deviate from the traditional picture of a sharp transition as formalised by \ct{zahn91}.
%To modify in addition the temperature gradient in the overshooting layer in another way than the Manichean approach of adopting either the radiative or adiabatic gradient, solar modellers need a more convincing argument.
% than just adding to the long list of parameter another one in order to improve the fit. 

The present work is motivated by arguments inspired by hydrodynamical simulations of convection and convective penetration in solar-like models. Recent hydrodynamical simulations by \ct[][hereafter B21]{baraffe21} highlight the process of 
 local heating in the overshooting region due to penetrating convective motions across the convective boundary. In the following, we analyse the potential impact of this feature on one-dimensional (1D) stellar evolution structures in the context of solar models. The hydrodynamical results of B21 are briefly summarised in Sect. \ref{hydro}, and their impact on 1D models are  analysed in Sect. \ref{1D} and discussed in Sect. \ref{conclusion}.

\section{Modification of the thermal background in the overshooting layer: Results from two-dimensional hydrodynamical simulations}
\label{hydro}
%\subsection{Results from two-dimension hydrodynamical simulations}
B21  performed two-dimensional (2D) fully compressible time-implicit simulations of convection and convective penetration in a solar-like model with the MUlti-dimensional Stellar Implicit Code MUSIC \cp[][]{viallet11, viallet16, goffrey17}. The main motivation was to explore the impact of an artificial increase in the stellar luminosity on the properties of convection and convective penetration. This procedure is a common tactic adopted in hydrodynamical simulations of convection \cp[][]{rogers06, meakin07, brun11, hotta17, edelmann19}. The experiments of B21 highlight the impact of 
penetrative downflows on the local thermal background in the overshooting layer.  They illustrate how convective downflows, when penetrating the region below the convective boundary of the envelope, can induce a local heating and a modification of the temperature gradient as a result of compression and shear in the overshooting layer. This modification of the local background is connected to a local increase in the radiative flux to counterbalance the negative enthalpy flux (or heat flux) produced by penetrating flows. The negative peak of the enthalpy flux and the positive bump of  the radiative flux  below the convective boundary are well-known features described in many numerical works  \cp[][]{hurlburt86, muthsam95, brummell02,brun11,  hotta17, kapyla19, cai20}. A few works \cp[][]{rogers06, viallet13, korre19, higl21} have also reported a 
modification of the local thermal background in the overshooting region, but without providing a detailed description. 
%This lack of deepening is most likely due to the fact that many of these simulations are either using an artificial increase of the luminosity or have not achieved thermal relaxation. 
The simulations of B21  provide a physical explanation that links the convective penetration process to the local heating and to the radiative bump in the overshooting layer. 
The solar-like star simulated in B21 is based on a model that is not thermally relaxed.  It is reasonable to assume  that the local heating seen in B21 is present in stars because the negative heat flux in the overshooting layer and the bump in the radiative flux that compensates for this feature are persistent.  These two features are also commonly observed in other hydrodynamical simulations, as mentioned above.  An exploration of the impact of this heating on stellar evolution models may reveal that heating is a necessary aspect of models for overshooting.
%Since these simulations are not thermally relaxed, the question is open whether this structural change can be applied to one-dimensional stellar models. But the negative heat flux in the overshooting layer and the bump in the radiative flux that compensates this feature are persistent and commonly observed in thermally relaxed simulations. It is thus reasonable to assume that this process of local heating described in B21 is present in stars and that it is worth exploring its impact on stellar evolution timescales, which is the purpose of this work.

\begin{figure}[h!]
%\vspace{-2cm}
\includegraphics[height=11cm,width=8cm]{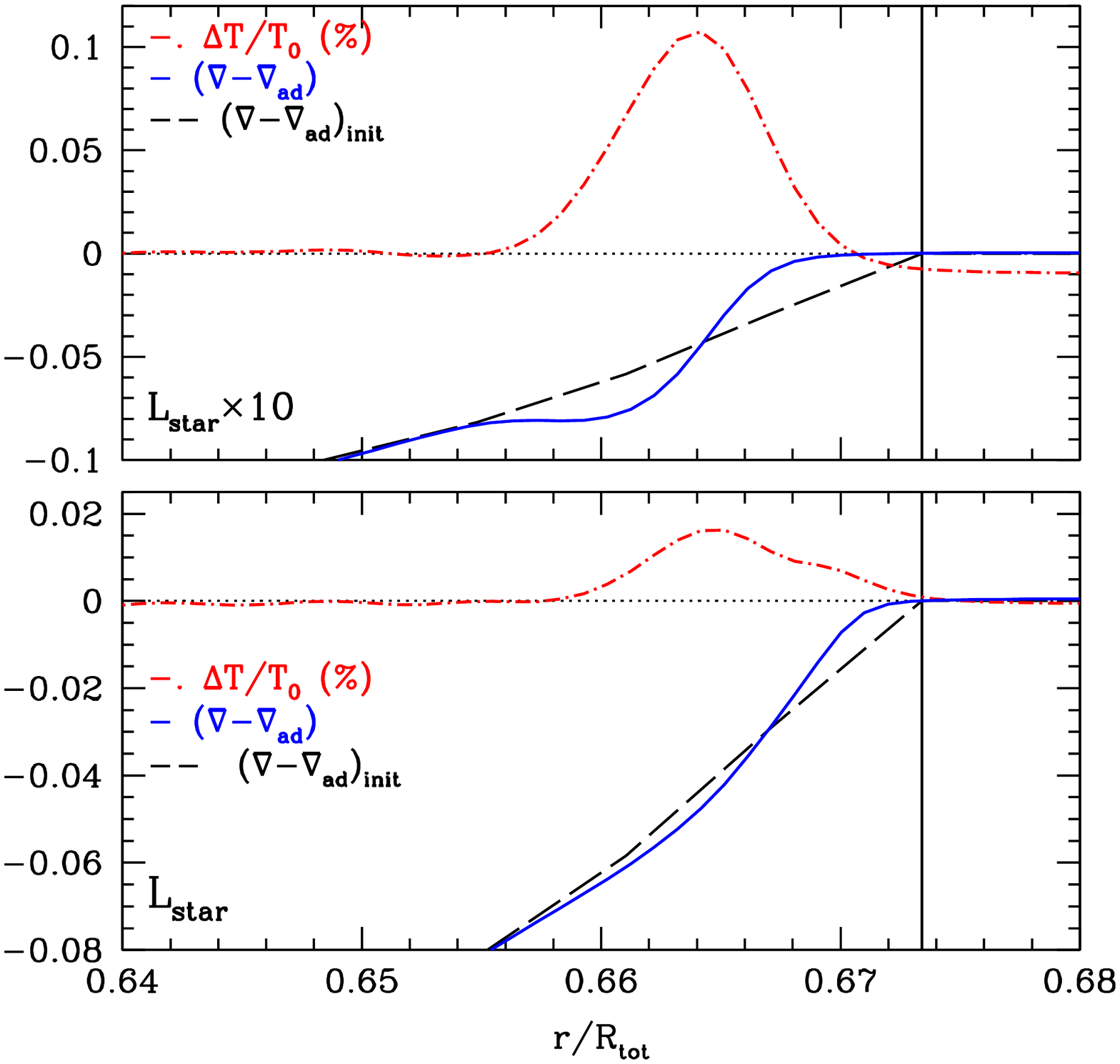}
%\vspace{-1.5cm}
   \caption{Radial profile of the temperature departure $\Delta T/T_0$ from the initial profile $T_0$ and of the sub-adiabaticity $(\nabla - \nabla_{\rm ad})$ close to the convective boundary predicted by 2D hydrodynamical simulations (B21) of solar-like models. The  lower panel corresponds to the model with a realistic stellar luminosity and the upper panel to a model with luminosity enhanced by a factor of ten.  The dash-dotted red lines show $\Delta T/T_0$  (in \%), the relative difference between the time and space averages of the temperature, $T$, and the initial temperature, $T_0$. The solid blue lines show the time and space averages of 
   the sub-adiabaticity $(\nabla - \nabla_{\rm ad})$. The dashed black lines show the initial profile of the
   sub-adiabaticity, $(\nabla - \nabla_{\rm ad})_{\rm init}$. The convective boundary is indicated by the vertical solid line (see details in B21)}
 \label{nabla2D_fig}
\end{figure}

The behaviour of the thermal profile below the convective boundary found in the simulations of B21 is illustrated in Fig. \ref{nabla2D_fig}. It is displayed for the model with a realistic stellar luminosity (lower panel). We also show  the results for a model with an artificial enhancement in the luminosity by a factor of ten because the features are intensified in these `boosted' models (upper panel). The figure shows 
the local heating in the overshooting layer  and its impact on the sub-adiabaticity  $(\nabla - \nabla_{\rm ad})$, with $\nabla = {{\rm d} \log T \over {\rm d} \log P}$ the temperature gradient and $\nabla_{\rm ad} = {{\rm d} \log T \over {\rm d} \log P}|_S$ the adiabatic gradient. The initial stratification below the convective boundary (located at $r=0.6734 \times R_{\rm star}$ for this specific stellar model) is set by the stable radiative gradient, $\nabla_{\rm rad}$  (see the dashed black line below the convective boundary in Fig. \ref{nabla2D_fig}). 
B21 show that, as a result of the local heating below the convective boundary characterised by the bump in temperature difference $\Delta T/T_0$ displayed  in Fig. \ref{nabla2D_fig}, the temperature gradient  becomes less sub-adiabatic immediately below the convective boundary{\footnote{ Less sub-adiabatic means that $|\nabla - \nabla_{\rm ad}|$ decreases compared to the initial profile.}.  %Just below that in terms of radius, the temperature gradient increases again with $|\nabla - \nabla_{\rm ad}|$ becoming larger than $|\nabla_{\rm rad} - \nabla_{\rm ad}|$. 
The net result is a smoother transition just below the convective boundary with a temperature gradient that has an intermediate value between the radiative temperature gradient and the adiabatic one. In the next section we analyse the impact of this local heating on 1D solar structures by adopting a simple prescription that mimics the behaviour of the temperature gradient suggested by hydrodynamical simulations.

\section{Impact on one-dimensional solar structure models}
\label{1D}

\subsection{Helioseismic constraints}
\label{seismology}

Our primary goal in this short paper is to illustrate the potential, qualitative impact of the local heating produced by overshooting.
%, rather than computing detailed solar models which will be the scope of a follow-up study.  
We adopted a
strategy inspired by the analysis of \ct{buldgen20},  who constructed a static structure of the Sun in agreement with seismic inversions of the Ledoux discriminant defined by
\begin{equation}
A = {1 \over \Gamma_1} {d \ln P \over d \ln r} - {d \ln \rho \over d \ln r },
\end{equation}
with $\Gamma_1 = (\partial \ln P / \partial \ln \rho)_{\rm ad}$.
Starting from a reference evolutionary model, \ct{buldgen20} used an inversion procedure to iteratively reconstruct a solar model. Successive inversions of the Ledoux discriminant allowed them to obtain a model-independent profile for this quantity. Their reconstruction method also gives solar structures that are in excellent agreement with other  structural inversions, namely the entropy, $S$, the square of the speed of sound, $c_{\rm s}^2$, and the density, $\rho$.
 To illustrate the convergence of their reconstruction procedure, they show (right panels of their Figs. 3-6) the successive iterations that converge to an excellent level of agreement for the four structural inversions ($A$, $S$, $c_{\rm s}^2$, $\rho$) starting from the initial reference model adopted in their work. The differences found between the reconstructed model and the reference model  are useful as they indicate the modifications of the reference model that are required to converge towards a solar model in agreement with  helioseismic data.  We recall here the major trends  found by \ct{buldgen20} for the four structural quantities, which are used for our analysis in Sect. \ref{test1d}.

The first concerns the Ledoux discriminant.\ The major discrepancy between the  Sun and the reference model occurs just below the convective boundary, with a large positive bump for the quantity ($A_{\rm Sun}$ - $A_{\rm ref}$). 

The second concerns the speed of sound. The same positive bump at the same location as for the Ledoux discriminant, $A$, is observed for the quantity $(c^2_{\rm s,Sun} - c^2_{\rm s,ref})/c^2_{\rm s,ref}$. The corrections applied to $A$ during the reconstruction procedure also reduce the discrepancy in the speed of sound in the radiative region.

The third concerns the entropy. Large discrepancies are observed in both the radiative region and the convective zone. The entropy discrepancy $(S_{\rm Sun}- S_{\rm ref})/S_{\rm ref}$ has two positive peaks in the radiative zone, one just below the overshooting region and a larger peak deeper at $\sim$ 40\% of the stellar radius. This discrepancy is  negative in the convective zone. The corrections applied to $A$ help reduce these entropy discrepancies in both regions.

The fourth concerns the density. The quantity $(\rho_{\rm Sun}-\rho_{\rm ref})/\rho_{\rm ref}$ has a negative peak in the radiative region, at $\sim$ 35\% of the stellar radius, and is positive in the convective zone. 

Importantly, \ct{buldgen20} mention that their reconstruction procedure gives  similar Ledoux discriminant profiles for a wide range of initial reference models. We used these results to gauge whether the modifications of the thermal profile  predicted by B21 can help in qualitatively improving all the structural quantities used by \ct{buldgen20}.

\subsection{Testing one-dimensional solar models}
\label{test1d}
Our main motivation is to show the potential  impact of the local heating 
described in Sect. \ref{hydro} on stellar models. We are not aiming in this short work at constructing the best solar model to fit helioseismic constraints.
Using stellar evolution codes, we have adopted two different methods that can be  found in the literature to construct solar models \cp[e.g.][]{zhang12, vinyoles17}.   
Our first method relies on the thermal relaxation of a reference model with solar radius and luminosity that is modified to reproduce the temperature gradient in the overshooting layer suggested by hydrodynamical simulations.  In this case, the chemical abundances are not modified by nuclear reactions, mixing, or microscopic diffusion during the relaxation process. For these tests, we used the 1D Lyon stellar evolution code \cp{baraffe98}. We repeated this experiment based on thermal relaxation with the stellar evolution code MONSTAR \cp[e.g.][]{constantino14} and obtained the same qualitative results.

The second method considers models that account for the modification of the temperature gradient in the overshooting layer from the zero age main sequence (ZAMS). The models are then evolved until they reach the solar radius and luminosity. With this approach, changes in the chemical abundances from nuclear reactions, microscopic diffusion, and overshooting mixing are  also consistent with any modification of the structure induced by the forced local heating in the overshooting layer. These tests 
 were performed with MONSTAR as it includes the treatment of microscopic diffusion.
 
 The first method allows the impact of local heating in the overshooting layer after thermal relaxation to be isolated. 
 The second method provides evolutionary models that are self-consistent since the effect of the modification of the temperature gradient is accounted for during their evolution on the main sequence. 
 %These two approaches can be used to construct a solar model with a stellar evolution code.

In the following, we adopt a modification of the local temperature gradient in the overshooting layer that qualitatively reproduces the behaviour displayed in Fig. \ref{nabla2D_fig}.
%The simplest assumption is to adopt a temperature gradient $\nabla$ between the radiative and adiabatic gradients; 
%\begin{equation}
%\nabla = {1 \over 2} (\nabla_{\rm ad} + \nabla_{\rm rad}). \label{eqnabla}
%\end{equation}
%\begin{eqnarray}
We define an overshooting length $d_{\rm ov} = \alpha_{\rm ov} H_{P,{\rm CB}}$, with $H_{P,{\rm CB}}$ the pressure scale height at the convective boundary and $\alpha_{\rm ov}$ a free parameter. We also define 
two radial locations, $r_{\rm ov} = r_{\rm CB} - d_{\rm ov}$ and $r_{\rm mid} = r_{\rm CB} - d_{\rm ov}/2$,
with $ r_{\rm CB}$ the radial location of the convective boundary. The temperature gradient is modified as follows. 
For  $r_{\rm mid} \le r < r_{\rm CB}$, we use
\begin{equation}
\nabla = g(r) \nabla_{\rm ad} + (1-g(r))\nabla_{\rm rad},    \label{eqnabla}
\end{equation}
with
\begin{equation}
g(r)=sin\{ [(r - r_{\rm mid})/(r_{\rm CB} - r_{\rm mid})]^a \times \pi/2\}. \label{eqg}
\end{equation}
%and we adopt $a$=0.3. 
For $r_{\rm ov} \le r < r_{\rm mid}$, we use 
\begin{equation}
\nabla = \nabla_{\rm rad} - h(r) \nabla_{\rm ad},
\end{equation}
with
\begin{equation}
h(r)=b \times sin\{ [(r_{\rm mid} -r)/(r_{\rm mid} - r_{\rm ov}) ] \times \pi \}.    \label{eqh}
\end{equation} 
Sine functions are used in Eqs. (\ref{eqg}) and (\ref{eqh}) to reproduce the smooth variations in the temperature gradient below the convective boundary produced by the hydrodynamical simulations.
We have verified that the results are insensitive to the smoothness of these variations and to the exact shape of the temperature gradient radial profile.We adopted $a$=0.3 in Eq. (\ref{eqg}) as it provides a behaviour for the temperature gradient very close to the one displayed in Fig. \ref{nabla2D_fig}. Results are rather insensitive to  variations in the values of $a$  between 0.2 and 0.4. We adopted $b$=0.03 in Eq. (\ref{eqh}), which also provides a close visual match to the hydrodynamical results, but we note that the results are  insensitive to the value of $b$.

%We have also verified that whether one uses smooth sinusoidal function or linear relations has no impact,
%as long as gradT remains rather close to grad_ad for r_mid <r < r_CB.

\subsubsection{Thermal equilibrium models}
\label{thermaleq}

The details of the procedure for the first method are the following.
%We perform a first series of tests with the one-dimensional Lyon stellar evolution code \cp{baraffe98} to analyse the general impact of the local heating described in \S \ref{hydro}. 
We calculate the evolution of  a 1 $\msol$ model   with an initial helium mass fraction of 0.28, metallicity $Z=0.02,$ and a mixing length $l_{\rm mix} = 1.9 H_P$. 
We use a reference model that is in thermal equilibrium\footnote{Thermal equilibrium means that the total nuclear energy produced in the central regions balances the radiative losses at the surface, {\it i.e.} the total nuclear luminosity, $L_{\rm nuc}$, equals the total stellar luminosity, $L$.} and has the luminosity and radius of the current Sun. 
%at an age of $\sim$ 4.7 Gyr. 
Starting from this reference model, the temperature gradient is modified over a prescribed depth to mimic the impact of overshooting according to the hydrodynamical simulations described in Sect. \ref{hydro}.
We adopt the prescription given by Eqs. (\ref{eqnabla})-(\ref{eqh}) over a distance $d_{\rm ov}$ below the convective boundary.
%with $H_{P,{\rm CB}}$ the pressure scale height at the convective boundary. %This value is in good agreement with the maximal depth reached by downflows below the convective boundary predicted by the hydrodynamical simulations for a solar like model. 
%We have varied this depth between 0.1$H_{P,{\rm CB}}$ and 0.2$H_{P,{\rm CB}}$, and the results do not change qualitatively.
We show the results in Fig. \ref{nabla1D_fig} for $\alpha_{\rm ov}$ = 0.15 and $\alpha_{\rm ov}$= 0.20. These overshooting widths are in good agreement with the maximal depth reached by downflows below the convective boundary predicted by the hydrodynamical simulations for the solar-like model investigated in B21. We note that the stellar model used in B21 is slightly under-luminous compared to the Sun (see B21 for details). B21 also mention that 
one should be cautious when directly applying the overshooting depths predicted by their simulations to real stars since the final relaxed state for these simulations may have different properties from non-thermally relaxed states. We varied $\alpha_{\rm ov}$ between 0.15 and 0.35 and find that the results do not change qualitatively. However, the amplitude of the variations in the model properties depends on  $d_{\rm ov}$ (see below).
As shown below, this simple prescription implemented in a stellar evolution code yields a local increase in the temperature below the convective boundary, similar to  that observed in the hydrodynamical simulations. We stress that  Eqs. (\ref{eqnabla})-(\ref{eqh}) have been chosen for simplicity. They are only a rough approximation that can mimic the thermal profile behaviour suggested in the 2D simulations.

 %Starting from this reference model, the temperature gradient is modified according to  Eq. (\ref{eqnabla}) over the prescribed depth. 
 The model with a modified temperature gradient is then thermally relaxed, that is to say, it is evolved over many thermal timescales without any modification of the abundances from nuclear reactions until thermal equilibrium is reached. The temperature gradient is modified in the overshooting layer during the whole relaxation process, and this is referred to as a `forced local heating'.
 This procedure ensures that the model with a modified temperature gradient can be consistently compared to the reference model. As shown in Fig. \ref{nabla1D_fig}, the simple prescription given by Eqs. (\ref{eqnabla})-(\ref{eqh}) yields similar qualitative changes in the temperature and the sub-adiabaticity  close to the convective boundary that was found in the hydrodynamical simulations of B21.

\begin{figure}[h!]
%\vspace{-1.5cm}
\includegraphics[height=13cm,width=8cm]{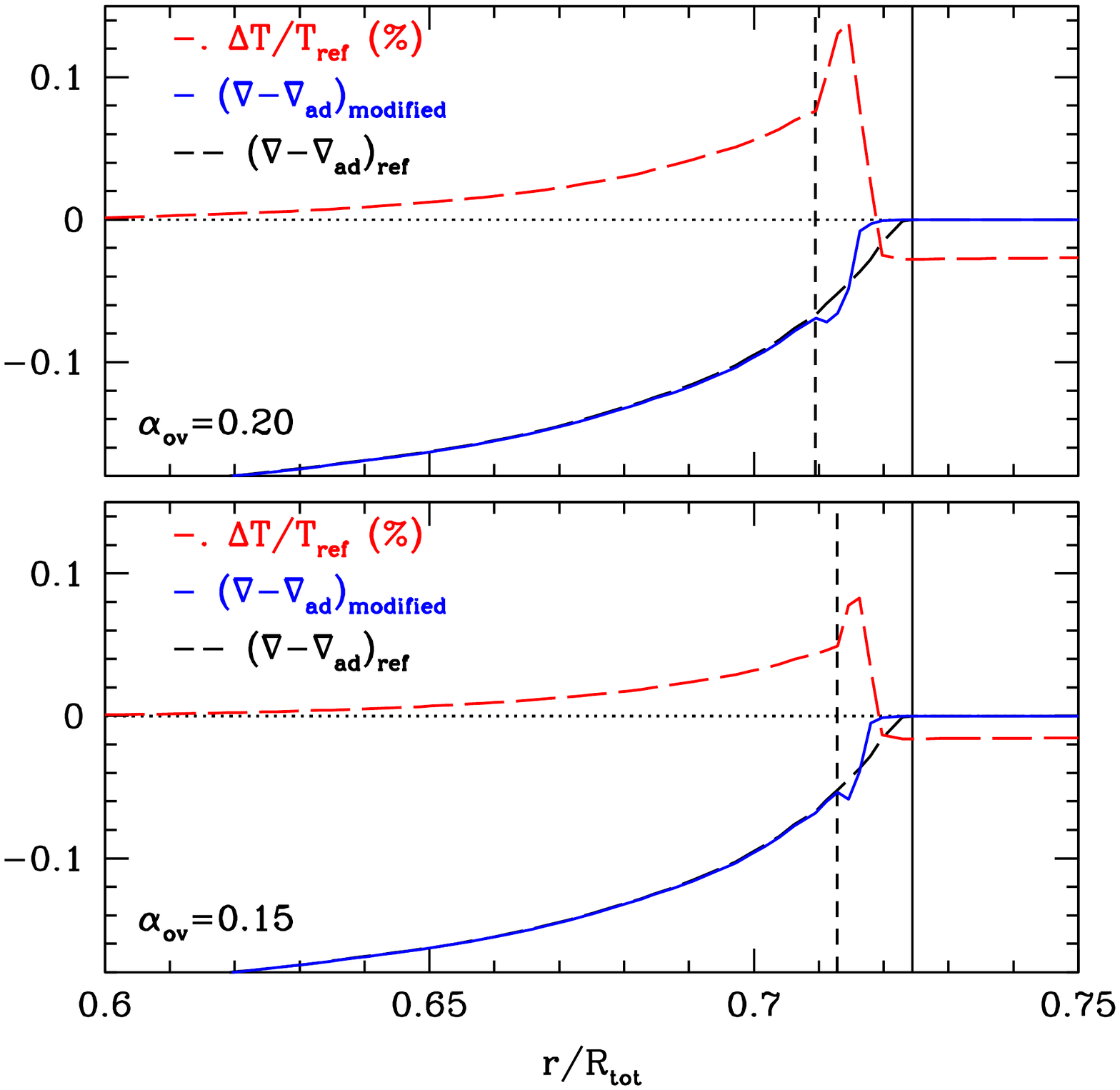}
%\vspace{-1.5cm}
   \caption{Radial profile of the temperature difference and of the sub-adiabaticity of a 1D solar-like structure with a modified temperature gradient in the overshooting layer according to Eqs. (\ref{eqnabla})-(\ref{eqh}). The temperature gradient is modified over a distance  $d_{\rm ov} = \alpha_{\rm ov} H_{P,{\rm CB}}$, with $\alpha_{\rm ov}$=0.15 in the lower panel and $\alpha_{\rm ov}$=0.20 in the upper panel.
  The dash-dotted red lines show the percentage relative temperature difference, $\Delta T / T_{\rm ref}$, with $\Delta T = T-T_{\rm ref}$. The solid blue lines correspond to the sub-adiabaticity $(\nabla - \nabla_{\rm ad})$. The dashed black lines show the sub-adiabaticity of the reference model. The convective boundary is indicated by the vertical solid line. The  vertical dashed line in each panel is located at a distance $d_{\rm ov}$ below the convective boundary. }
 \label{nabla1D_fig}
\end{figure}

The impact on the whole stellar structure was quantified by comparing the four structural quantities ($A$, $S$, $c_{\rm s}^2$, $\rho$) between the modified and the reference model. 
The results are displayed in Fig. \ref{diff_fig}, with $\Delta X$ defined as $(X - X_{\rm ref}$) for any structural quantity $X$. The forced local heating in the overshooting layer produces similar positive peaks for $\Delta A$, $\Delta S$, and $\Delta c_{\rm s}^2$, as found for the temperature.  The modification thus provides the correction required to improve the discrepancy for the Ledoux discriminant described 
in the first of the trends outlined in Sect. \ref{seismology}.
Unsurprisingly, such a modification of the temperature gradient is expected to improve the agreement with helioseismic constraints and help remove the sound speed anomaly below the convective boundary (second trend in Sect. \ref{seismology}), as suggested by the results of \ct{christensen11}. But it is also interesting to note that such a modification yields a slight cooling of the convective zone (see Fig. \ref{nabla1D_fig}) and thus a negative difference for the entropy (see  Fig. \ref{diff_fig}).  
A negative difference in the convective envelope is in agreement with the correction required for the reference model of \ct{buldgen20} to better match the Sun (see third trend in Sect. \ref{seismology}).  
Regarding the density, the modification of the temperature gradient has an interesting impact in the radiative zone, with a large decrease in the density compared to the reference model over a broad region below the convective boundary. The impact on the density in the convective region for this specific model is  partly in agreement with the correction required for this quantity in the \ct{buldgen20} study,  with a positive difference found only in the upper part of the convective envelope} (see the fourth trend in Sect. \ref{seismology}).

These trends are insensitive to the depth over which the temperature gradient is modified. Increasing  the depth increases the magnitude of the differences but has no impact on their sign.  { We find that the maximum variation in the model properties, such as the speed of sound, $\Delta c^2_{\rm s} /c^2_{\rm s,ref}$, roughly scales with $d_{\rm ov}^2$. This scaling is linked  to the integrated area between the modified temperature gradient curve and the one for the reference (non-modified) temperature gradient, which roughly decreases linearly with $r$. This area is proportional to the square of the overshooting depth, and consequently, the maximum variation in the model properties is also proportional to $d_{\rm ov}^2$.
%The trends are also insensitive to the treatment of chemical mixing in the overshooting layer. 
The qualitative trends also remain the same whether overshooting mixing in the reference model is ignored or included using  a step function (with instantaneous mixing) or an exponential decay for the diffusion coefficient \cp[e.g.][]{freytag96}.

 %with a does not seem to help reducing the discrepancy between $\rho_{\rm Sun}$ and $\rho_{\rm ref}$ of the \ct{buldgen20} study. 

\begin{figure}[h!]
%\vspace{-2cm}
\includegraphics[height=14cm,width=8cm]{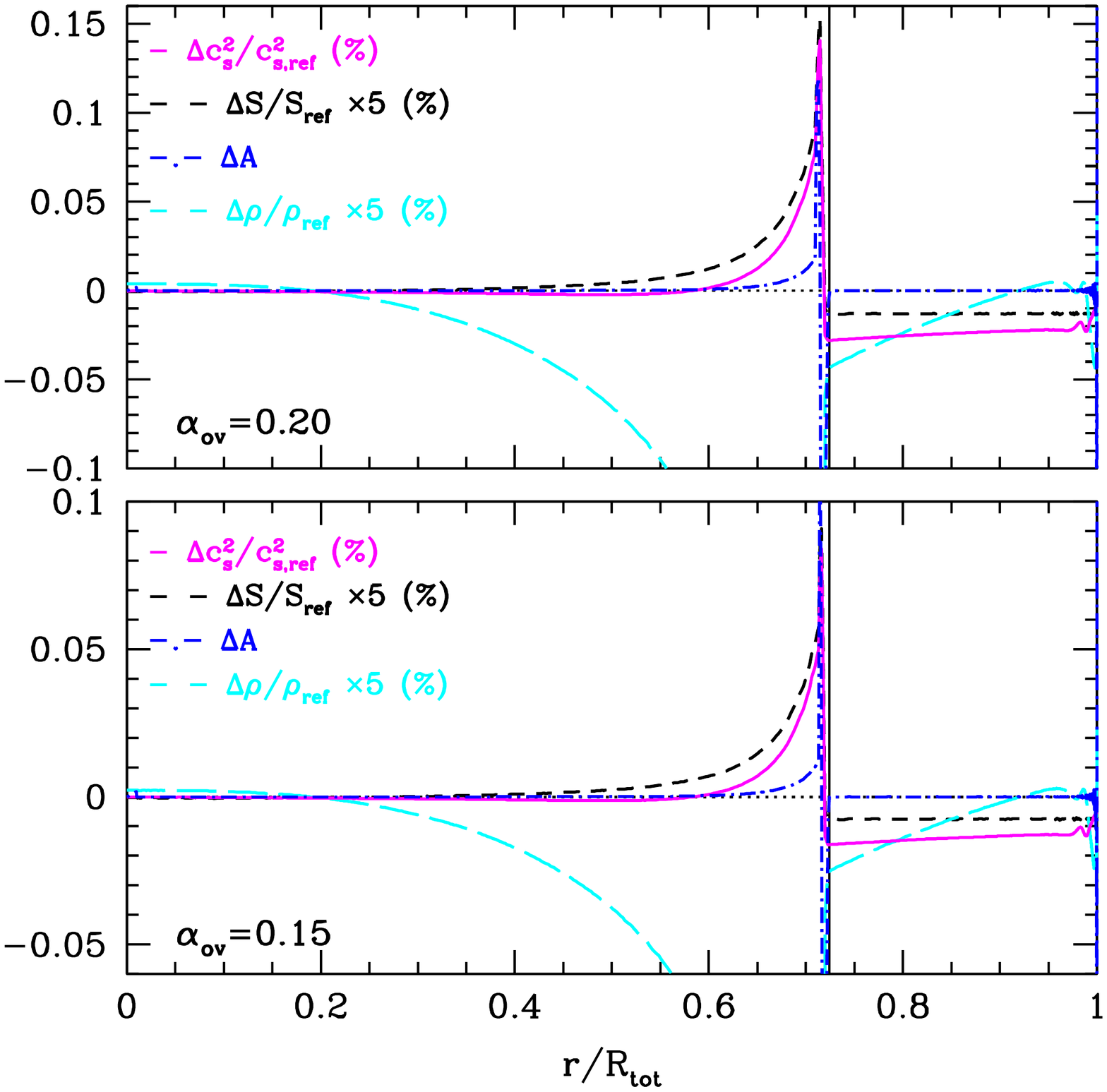}
%\vspace{-2cm}
   \caption{Difference of various structural quantities between a model with a modified temperature gradient in the overshooting layer and a reference model calculated with the Lyon stellar evolution code. The  temperature gradient in the modified model is changed over a distance $d_{\rm ov} = \alpha_{\rm ov} H_{P,{\rm CB}}$ below the convective boundary (indicated by the vertical solid line). The lower panel shows the results for  $\alpha_{\rm ov}=0.15$ and the upper panel for $\alpha_{\rm ov}=0.20$.}
 \label{diff_fig}
\end{figure}

%To test the robustness of the trends found for the four structural quantities, we have also performed tests with two other stellar evolution codes. The second series of tests are performed with the MESA code \cp{paxton11}. 
%which includes microscopic diffusion \cp{paquette86, thoul94}. The later process is not included in the Lyon stellar evolution code. 
%We adopt the same strategy as done with the Lyon code, namely modifications are performed starting from a relaxed reference model with the solar radius and luminosity. The tests with MESA confirm the effects of the modification of the temperature gradient found with the Lyon code for the four structural quantities. 
%ests with microscopic diffusion....TBD

\subsubsection{Self-consistent evolutionary models}
\label{self}

For the tests based on the second method, 
 %are performed with the stellar evolution code MONSTAR \cp[e.g.][]{constantino14}, as it includes the treatment of microscopic diffusion.
 we ran different sets of models with different combinations of assumptions, including or not microscopic diffusion and with or without overshooting mixing. 
When overshooting mixing was included in the overshooting layer, it was based either on a step function  or on an exponential decay for the diffusion coefficient. Microscopic diffusion for H and He was implemented according to \ct{thoul94}. For these tests, the temperature gradient was modified  according to { Eqs. (\ref{eqnabla})-(\ref{eqh})}. All models start from the ZAMS and are evolved until they reach the solar radius and luminosity at the same age. This was achieved by making small adjustments to the mixing length,  $l_{\rm mix}$. The models with temperature gradient modifications were compared to the relevant reference model, which has no modification of the temperature gradient but everything else is the same (i.e. the same treatment of microscopic diffusion and of overshooting mixing).  
The evolutionary models with temperature gradient modifications are thus self-consistent.
%as the effect of the modification of the temperature gradient is accounted for during their evolution on the main sequence. 
The main difference between this approach and the one in the previous section is  that these models accumulate small
differences in, for example, central H abundance when compared to their reference model. These tests produce the same trends  in the overshooting layer as found for the tests based on the first method (Sect. \ref{thermaleq}), independently of the treatment of overshooting mixing and whether microscopic diffusion is included or not. In the convective zone, all models give a positive difference for the density  between the model with a modified temperature gradient and the relevant reference model.
For the other quantities ($S$, $c_{\rm s}^2$), the differences in the convective zone are very sensitive to the assumptions regarding whether overshooting mixing is included or not. But at least we find solutions that are compatible with the four trends found by \ct{buldgen20} for the four structural quantities. This is illustrated in  Fig. \ref{monstar_fig} with a model that accounts for step function overshooting mixing  over a distance $d_{\rm ov} = 0.15 H_{P,{\rm CB}}$ (lower panel) and $d_{\rm ov} = 0.20 H_{P,{\rm CB}}$ (upper panel).
 
 \begin{figure}[h!]
%\vspace{-2.5cm}
\includegraphics[height=14cm,width=8cm]{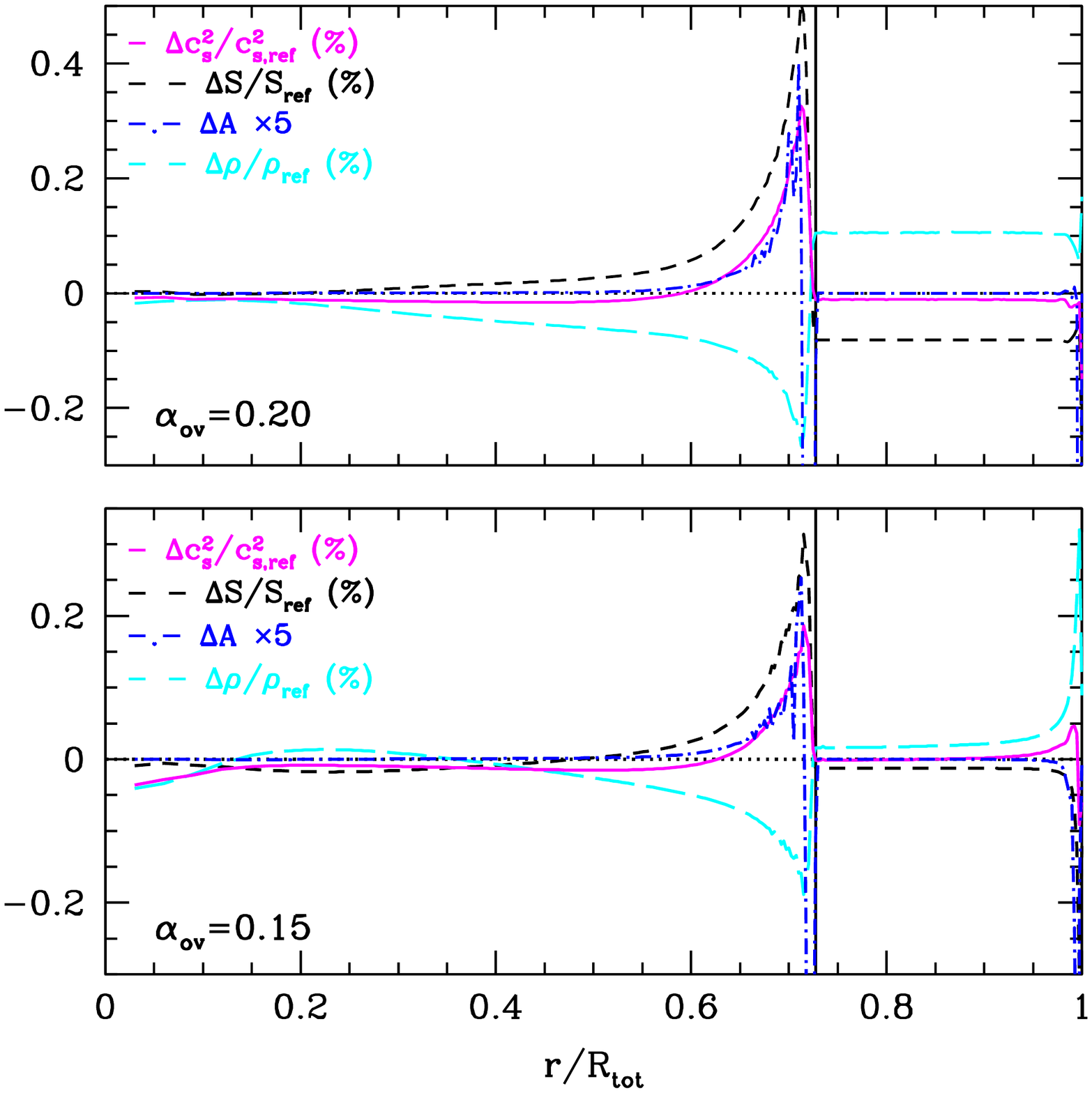}
%\vspace{-2cm}
   \caption{Difference of various structural quantities between a modified model and a reference model calculated with the MONSTAR stellar evolution code. The reference model is evolved from the ZAMS with microscopic diffusion and step function overshooting mixing over a distance $d_{\rm ov}= \alpha_{\rm ov} H_{P,{\rm CB}}$ below the convective boundary. {The lower panel shows the results for  $\alpha_{\rm ov}=0.15$ and the upper panel for $\alpha_{\rm ov}=0.20$.} The models with a modified temperature gradient in the overshooting layer (same microscopic diffusion and overshooting mixing treatment as the reference model)  are evolved similarly from the ZAMS. The convective boundary is indicated by the vertical solid line.}
 \label{monstar_fig}
\end{figure}

\section{Conclusion}
\label{conclusion}
The tests performed in Sect. \ref{1D} are based on different methods (relaxed models versus consistent evolution) that can be used to construct solar models. Independently of the method used, the tests show
 that a local increase in the temperature in the overshooting region due to convective penetration provides the qualitative effects  required to improve the speed of sound discrepancy below the convective boundary. This discrepancy is persistent in solar models that use low solar metal abundances.  This is not surprising because 
an increase in the temperature in this specific region has previously been invoked in the literature to solve this problem, as mentioned in Sect. \ref{introduction}. However, the details of the physical process responsible for this local heating have been lacking, whereas we can now suggest an explanation based on the B21 results. The trends that we find for the four structural quantities ($A$, $S$, $c_{\rm s}^2$, $\rho$) are robust below the convective boundary and in a large fraction of the radiative core, independently of the treatment of mixing and diffusion and of the method for constructing the models in Sects. \ref{thermaleq} and \ref{self}.
Our experiments additionally show that such a local change in the temperature, despite being made over a very limited region below the convective boundary, can also affect the density, the entropy, and the speed of sound in the convective envelope after thermal relaxation or evolution on the main sequence. 
How these quantities are affected in the convective envelope compared to  a reference model with no local heating 
depends on the strategy for building solar models and on the treatment of overshooting mixing. This mixing is obviously  linked to the
local heating given that both result from the same dynamical process. A combined testing of both effects in stellar models could provide more constraints on the general process of overshooting.  

Increasingly, efforts are now devoted to characterising the process of convective boundary mixing in stellar models based on multi-dimensional hydrodynamical simulations. More work is required  to obtain reliable determinations of an overshooting depth and to describe quantitatively the mixing and impact on the temperature gradient.  Understanding the effects of rotation and magnetic fields on overshooting is a significantly more difficult theoretical and numerical problem to address; however, efforts to study these combined non-linear effects are ongoing \cp{hotta17, korre21}.
Despite the limitations of existing hydrodynamical simulations, they are already providing constraints on physical processes usually treated with several free parameters in 1D stellar evolution models. They can thus limit the degrees of freedom in a problem as complex as solar modelling. Our primary goal in this work is to highlight 
the potential impact of convective penetration on the thermal background in the overshooting region. The processes studied in B21 that produce  a local change in the temperature gradient are also responsible for the mixing in this region. Because much observational evidence points towards the need for extra mixing at convective boundaries, for example lithium depletion in solar-like stars \cp{baraffe17}, the size of convective cores \cp{claret16}, and colour-magnitude diagrams \cp{castro14}, solar modellers often include this extra mixing in their models. But a consistent approach should also require accounting for a local change in the temperature gradient. The impact of this local heating goes in the right direction to improve not only the discrepancies of solar models below the convective boundary, but also in the convective envelope. This effect offers an interesting step forward for solving the solar modelling problem. 
In this exploratory work, we adopt a simple prescription for the local heating in the overshooting layer since the main goal is to highlight its qualitative impact on stellar models. However, this effect should not be considered as another free parameter in the solar modelling problem. Future multi-dimensional hydrodynamical simulations will enable this process, and its treatment in 1D stellar evolution codes, to be better constrained. 
 
%We also plan to investigate the impact of this effect on more detailed and realistic solar models to compare with helioseismic data and on lithium depletion, which provides a powerful complementary constraint on the process of overshooting. 

%I think it is worth pointing out that changing gradT does affect helioseismic comparisons and that this may be justified by hydro simulations + theory (in the latter case there is at least the potential for a contradiction if composition can mix but the heat is ignored).  It's therefore another thing along with composition, opacity etc. they should explore in their models.  I would guess that there is always going to be some flexibility in "solving" the solar problem because there are a few variables to play around with.
%The best angle of attack, as you say, is to highlight the importance of changing gradT, rather than convincing that this would solve the solar problem... I am even not convinced myself, given that there are indeed many other parameters, but I think it is important that people realise that it is certainly more physical to change gradT than not to change it.

 \section{Acknowledgements}
We thank our anonymous referee for valuable comments which helped improving the manuscript. This work is supported by the ERC grant No. 787361-COBOM and the consolidated STFC grant ST/R000395/1. IB thanks the Max Planck Institut f{\"u}r Astrophysics (Garching) for warm hospitality during completion of part of this work. The authors would like to acknowledge the use of the University of Exeter High-Performance Computing (HPC) facility ISCA and of the DiRAC Data Intensive service at Leicester, operated by the University of Leicester IT Services, which forms part of the STFC DiRAC HPC Facility. The equipment was funded by BEIS capital funding via STFC capital grants ST/K000373/1 and ST/R002363/1 and STFC DiRAC Operations grant ST/R001014/1. DiRAC is part of the National e-Infrastructure.

\bibliographystyle{aa}
\bibliography{references}

\end{document}